\documentclass[eqsecnum,superscriptaddress,showpacs]{revtex4}
\usepackage{graphics}
\usepackage{latexsym}
\usepackage{bm}
\begin{document}
\title{Tomonaga's Conjecture on Photon Self-Energy}
\author{Takehisa Fujita}\email{fffujita@phys.cst.nihon-u.ac.jp}
\author{Naohiro Kanda}\email{nkanda@phys.cst.nihon-u.ac.jp}
\affiliation{Department of Physics, Faculty of Science and Technology, 
Nihon University, Tokyo, Japan}


\begin{abstract}
We investigate Tomonaga's conjecture that the self-energy of photon should vanish 
to zero.  In fact, the contribution of photon's self-energy diagram violates the Lorentz 
invariance and therefore it is unphysical. In addition, there occurs no wave 
function renormalization of photon in the exact Lippmann-Schwinger equation 
for the vector potential and this confirms that the conjecture is correct. 
Further, it is shown that the gauge condition $ k_\mu \Pi^{\mu \nu}(k) =0$ 
of the vacuum polarization tensor does not hold and the relation is obtained simply  
due to a mathematical mistake in replacing the integration variables in the infinite 
integral. 

\end{abstract}


\maketitle

\maketitle

\section{Introduction}
A half century ago, Tomonaga stated that the self-energy of photon should vanish 
to zero even though he did not present any concrete proof at that time \cite{tomonagac}. 
Since then, however, people have never considered his conjecture seriously, and up to now, 
the self-energy of photon has been included into the renormalization scheme by throwing 
away the quadratic divergence terms and by keeping only the terms which have the same shape 
as the Lagrangian density of photon. In his statement, he also claimed that 
the regularization scheme proposed by Pauli and Villars \cite{pauli} should not be a proper 
method for the renormalization scheme, but unfortunately, however, there was no concrete 
proof to this statement either, as far as we can check his published works as 
research papers. 

In this paper, we study his conjecture that the self-energy of photon should vanish 
to zero and prove that it is indeed a correct statement. The basic reason 
why the self-energy of photon should vanish is basically because the self-energy of photon 
arising from the vacuum polarization diagram violates the Lorentz invariance and 
therefore the process of the photon self-energy diagram is unphysical, and this is 
similar to the Feynman diagrams which violate the energy and momentum conservations. 
Further, the exact Lippmann-Schwinger equation for the vector potential $\bm{A}$ 
clearly shows  that the vector field cannot be affected from the perturbation expansion 
\cite{lipp}. Since this is the exact equation of motion for  $\bm{A}$, there is no way 
that the vector potential can be influenced by any of the vacuum polarization effects. 

The basic physics picture of the vacuum polarization was made up by Heisenberg in 1934, 
and in order to clarify the problem of the vacuum polarization effects, we are forced to 
examining the papers by Heisenberg and Euler \cite{heisen,heisen2}. 
The evaluation of the self-energy 
of photon was carried out just in a similar way to the Feynman diagram calculations. 
They obtained the quadratic divergence and logarithmic divergence terms, in addition 
to the finite effects of the vacuum polarization, and they considered the corresponding 
Lagrangian density. The effective Lagrangian density Heisenberg and Euler proposed can 
take into account the effects of the vacuum polarization. Apart from the renormalization 
procedure, the vacuum polarization effect is almost the same as that of the modern version. 
However, if one looks into their calculation carefully, then one can easily realize that 
the effective Lagrangian density method is a dangerous attempt for physics. 
The Lagrangian density is the basic tool which can describe the physical processes, 
and if one changes its shape, then this means that the basic physical law itself is 
modified even if the matter fields are not present. This is indeed connected to 
the understanding of the concept of the vacuum in field theory. In fact, the fields 
which appear in the effective Lagrangian density should become operators in the next 
step calculations, and this should produce unphysical processes as a result. 

On the other hand, the renormalization scheme of the self-energy of photon in the present 
day is to find and determine the same shape of the Lagrangian density after 
the renormalization of the vacuum polarization contributions. However, the procedure 
of the renormalization scheme of the self-energy of photon is entirely based on 
one simple relation which is called "gauge condition" for the vacuum polarization tensor. 
In the renormalization scheme of QED, this gauge condition played the most important role 
as the excuse for throwing away the quadratic divergence contribution. Here, we show 
in a simple mathematical fashion that the gauge condition of  $\Pi^{\mu \nu}(k)$ 
$$ k_\mu \Pi^{\mu \nu}(k) =0   \eqno{(1.1)}  $$
is invalid. As we show later, the proof of eq.(1.1) which has been done in most of 
the field theory textbooks is based on a mathematical mistake which is basically 
due to the wrong replacement of the integration variable in the infinite integral 
case. This problem of mathematics was, of course, realized and stated in the text 
book of Bjorken and Drell \cite{bd}, but they made the same type of mistake in 
evaluating the vacuum polarization contributions in the regularization scheme 
by Pauli and Villars \cite{pauli}. In this respect, one sees that the Pauli-Villars 
regularization does not change anything in physics, and it is a meaningless procedure. 
This can be well realized if one calculates any physical processes of loop diagrams 
such as $\pi^0 \rightarrow 2 \gamma$ and the box diagram in photo-photon scattering, 
and they have no logarithmic divergences. Indeed, the evaluation of all of these 
physical processes shows that they are finite due to the kinematical cancellation 
of the divergent contributions, and in fact the Pauli and Villars regularization 
does not play any role in these calculations of the Feynman diagrams.

Before going to the discussion of the gauge condition of eq.(1.1), we should 
understand the physical meaning of the gauge invariance in connection with 
the QED renormalization scheme \cite{tomonaga,tomonaga2,schw,schw2,feyn}. 
Any physical quantities calculated by fixing the gauge should not depend on 
the gauge choice, and in fact the gauge invariance of the electromagnetic interaction 
$$ H'= -e\int j_\mu A^\mu d^3r  \eqno{(1.2)}  $$ 
can be guaranteed with the condition that the fermion current should be conserved, 
that is 
$$ \partial_\mu j^\mu=0 .  \eqno{(1.3)}  $$ 
The gauge invariance of $H'$ can be easily seen under the gauge 
transformation of $A_\mu ={A'}_\mu+\partial_\mu \chi$ as
$$ H'=-e\int j_\mu ({A}^\mu-\partial^\mu \chi) d^3r
=-e\int j_\mu {A}^\mu d^3r+e\int \partial_\mu (j^\mu \chi) d^3r 
=-e\int j_\mu {A}^\mu d^3r   \eqno{(1.4)}  $$ 
where the surface integral vanishes and we made use of eq.(1.3). 
This is the most important condition of the gauge invariance in QED, and as long as 
we carry out the perturbation theory with free fermion basis states, 
then the calculated results are always gauge invariant. There is no additional 
constraint like eq.(1.1) in the QED renormalization scheme \cite{fujita}. 
Since the calculations with eq.(1.1) give wrong answers to the evaluation of 
many basic Feynman diagrams, it is important to clarify the situation 
at the present stage. 

\section{Tomonaga's Conjecture}
In this section, we prove that the self-energy of photon should vanish 
to zero. First, we start from the QED Lagrangian density which can describe 
the fermions interacting with the electromagnetic fields. It reads
$$ {\cal L} = \bar{\psi} (i\partial_\mu \gamma^\mu -gA_\mu \gamma^\mu-m)\psi 
- {1\over 4}F_{\mu \nu}F^{\mu \nu} 
\eqno{(2.1)} $$
where $ F_{\mu \nu} $ denotes the field strength and is given as
$$ {F}_{\mu \nu}=\partial_\mu {A}_\nu-\partial_\nu {A}_\mu . \eqno{(2.2)} $$
$ A^\mu$ denotes the gauge field with $  A^\mu =(A_0, \bm{A}) $.  
In this case, we can obtain the Dirac equation with the electromagnetic interaction 
$$ (i\partial_\mu \gamma^\mu -gA_\mu \gamma^\mu-m)\psi =0. \eqno{(2.3)}  $$
Also, we find the Maxwell equation
$$ \partial_\mu F^{\mu \nu} =g j^\nu \eqno{(2.4)} $$
where the current density $j^\nu$ is defined as 
$$ j^\nu =\bar{\psi} \gamma^\nu \psi  . \eqno{(2.5)} $$
As the gauge fixing condition, we can take the Coulomb gauge 
$$ \bm{\nabla}\cdot \bm{A} =0 . $$

\subsection{Lippmann-Schwinger equations}
If one wishes to carry out the S-matrix evaluation, then the Lippmann-Schwinger equations 
must be most suitable for figuring out what are physical observables out of the S-matrix 
contributions \cite{lipp}. Therefore, we first write the Lippmann-Schwinger equations 
for fermions 
and vector field. The Lippmann-Schwinger equation for the fermion field $\psi$ becomes 
$$ \psi (x) = \psi_0(x)+  g \int G_F(x,x') A_\mu(x') \gamma^\mu \psi(x') d^4x' 
\eqno{(2.6)}  $$
where $G_F(x,x')$ denotes the Green function which satisfies the following equation
$$ (i\partial_\mu \gamma^\mu -m) G_F(x,x') =\delta^4 (x-x'). \eqno{(2.7)}  $$
$\psi_0(x)$ is the free fermion field solution. The Green function $G_F(x,x')$ can be 
explicitly written as
$$ G_F (x,x') = \int {1\over p_\mu \gamma^\mu -m+i\epsilon} e^{ip(x-x')} 
 {d^4p\over(2\pi)^4} . \eqno{(2.8)}  $$
On the other hand, the Lippmann-Schwinger equation for the vector field $\bm{A}$  becomes 
$$ \bm{A} =  \bm{A}_0 +  g \int G_0(x,x') \bm{j}(x') d^4x' \eqno{(2.9)}  $$
where $ \bm{A}_0 $ denotes the free field solution of the vector field. Here, the Green 
function $G_0(x,x') $ satisfies 
$$ \left({\partial^2\over \partial t^2}-\bm{\nabla}^2\right) G_0(x,x') =\delta^4 (x-x'). 
\eqno{(2.10)}  $$
This can be explicitly written as
$$ G_0 (x,x') = \int {1\over -p_0^2+\bm{p}^2 +i\epsilon} e^{ip(x-x')} 
 {d^4p\over(2\pi)^4} . \eqno{(2.11)}  $$
Since we employ the Coulomb gauge fixing, the equation of motion for the $A_0$ field 
becomes a constraint equation and thus can be solved exactly as
$$ A_0 (\bm{r}) ={g\over 4\pi} \int {j_0(\bm{r}') \over |\bm{r}-\bm{r}'| } d^3r' . 
 \eqno{(2.12)}  $$
This is all that we need for the discussion of the renormalization procedure. 
Up to this point, we have not made any approximations for the equations of motion. 
Now we can quantize the fermion field $\psi$ as well as the vector field $\bm{A}$, and 
the quantized fields should be denoted as $  \hat{\psi}, \ \hat{\bm{A}}$. 

\subsection{Wave Function Renormalization$-$Fermion Field}
When  we carry out the perturbation expansion, we can obtain the integral equations 
in powers of the coupling constant $g$ as
$$ \hat{\psi} (x) = \hat{\psi}_0(x)+  g \int G_F(x,x') \hat{A}_\mu \gamma^\mu 
\hat{\psi}_0(x') d^4x' +
g^2 \int G_F(x,x') \hat{A}_\mu(x') \gamma^\mu G_F(x',x'') \hat{A}_\nu(x'') \gamma^\nu 
\hat{\psi}_0(x'')  d^4x'd^4x''+\cdots . \eqno{(2.13)}  $$
This equation clearly shows that the fermion field should be affected by 
the perturbation expansion, and if it diverges, then we have to renormalize the wave 
function so as to absorb the infinity. Indeed, the infinity is logarithmic divergence 
and can be well renormalized into the wave function $\psi_0$. 

\subsection{Wave Function Renormalization$-$Vector Field}
On the other hand, the situation for the vector field case is quite different 
from the fermion field. This can be easily seen from the equation of motion of eq.(2.9). 
In this case, there is no iteration possible for the $\bm{A}$ field. The vector field  
$\hat{\bm{A}}$ can be determined from eq.(2.9) only when the fermion numbers are conserved. 
The best example can be found when the annihilation of the fermion pair takes place. 
In this case, we can write eq. (2.9) as
$$ \langle 0| \bm{A}|f \bar{f}\rangle  = g \int G_0(x,x') 
\langle 0|  \bm{j}(x') |f \bar{f}\rangle d^4x' \eqno{(2.14)}  $$
where $|f \bar{f}\rangle $ denotes the fermion and anti-fermion state. 
Now, we can consider the following physical process of $e^+e^- \rightarrow  e^+e^- $  
which can be described in terms of the T-matrix as
$$ T = -g \langle e \bar{e}| \int \bm{j}\cdot \bm{A} |e \bar{e}\rangle 
=-g^2 \int \langle e \bar{e}| \bm{j}(x)|0\rangle G_0(x,x') \langle 0| \bm{j}(x') 
|e \bar{e}\rangle d^3x d^4x'  \eqno{(2.15)}  $$
where one can see that there appears no self-energy of photon term whatever one evaluates 
any physical processes in the Lippmann-Schwinger equation. 
From this equation,  one finds that the vector field $\hat{\bm{A}}$ cannot be 
affected by the renormalization procedure, and it always stays as a free state 
of photon. Since this is the exact equation of motion, there is no other possibility 
for the vector field. In this respect, it is just simple that the gauge field 
 $\hat{\bm{A}}$  always behaves as a free photon state in the evaluation 
of any Feynman diagrams. 

\subsection{Mass Renormalization}
In the evaluation of the S-matrix in QED, there are some Feynman diagrams which 
are divergent, and therefore we should renormalize them into the mass term if 
they are consistent with fundamental symmetries. 

\subsubsection{Fermion Self-energy}
The evaluation of the self-energy of fermions can be carried out in a straight 
forward way, and one can obtain the self-energy which has a logarithmic divergence 
of the momentum cut-off $\Lambda$. Since electron has a mass, one can renormalize this 
logarithmic divergence term into the new mass term. In this procedure, there is 
no conceptual difficulty and indeed one can relate this renormalized effect 
to the observed value of the Lamb shift in hydrogen atom, which is indeed a great 
success of the QED renormalization scheme. 

Here, we should note that the evaluation of the Lamb shift energy has been done only 
when the non-relativistic wave function of the hydrogen atom is employed. 
This is made by Bethe \cite{bethe}, and he stated that the relativistic treatment should be 
in progress. However, since then, there has been no work presented which treats 
the relativistic wave function of hydrogen atom. This is, however, quite understandable 
since we cannot calculate the negative energy states of the hydrogen atom up to the present 
stage. On the other hand, people believe that they have calculated the Lamb shift energy 
relativistically, but this is only the relativistic  correction on the Lamb 
shift energy, and the main term of the Lamb shift energy is still far from satisfactory 
since the Bethe's treatment has the logarithmic divergence of the Lamb shift energy. 
But, in other words, since it is logarithmic divergence, it should be reliable 
for the Lamb shift energy up to the order of the magnitude evaluation.

\subsubsection{Photon Self-energy}
On the other hand, the evaluation of the self-energy of photon gives rise to 
the energy which has a quadratic divergence. There is no way to renormalize it 
into the renormalization scheme of QED since photon has no mass term. This clearly 
indicates that one should not consider the contributions of the photon self-energy 
diagrams into the renormalization scheme since they violate the Lorentz invariance. 
In this respect, one can now realize that the quadratic divergence term should be 
discarded because it is not consistent with the Lorentz invariance, and it has nothing 
to do with the gauge invariance. 

For the energy of photon, we should be careful in which system we are calculating it. 
This is very important since, obviously, there is no system where photon is at rest. 
The system we are calculating should be the system in which fermion is at rest. 
In this system, the energy of photon with its momentum $\bm{k}$ must be described 
as $ E_k=|\bm{k}|$, and there is no other expression. Therefore, the Lagrangian density 
of the vector field $A_\mu$ should be always written as
$$ {\cal L}_0=-{1\over 4} F_{\mu \nu} F^{\mu \nu} \eqno{(2.16)} $$
and there should not be any modifications possible. 

\section{Vacuum Polarization- Revisit}
Up to this point, we have clarified that the self-energy of photon should vanish 
to zero, and therefore photon stays always massless and there is no effect from 
the renormalization procedure. However, there have been many works for the QED evaluation 
with the renormalization procedure which considered the renormalization of the vacuum 
polarization contribution. Therefore, we should clarify what should be the basic mistake 
in the treatment commonly used up to the present stage. 

First, we should like to critically review the renormalization procedure of 
the self-energy of photon. The basic starting point of the renormalization 
procedure is connected to the quantity of the vacuum polarization tensor 
$\Pi^{\mu \nu}(k)$. This quantity itself is not directly a physical observable, 
but somehow people have been making use of the vacuum polarization tensor. 

\subsection{Vacuum Polarization Tensor}
If one carries out the self-energy diagram of photon, then one obtains 
the quadratic divergence contributions. This gives rise to some difficult 
problems, and here we show that there is no way to renormalize the contributions 
of the self-energy of photon into the Lagrangian density as the counter terms.  
First, we write the result of the standard calculation of the vacuum polarization 
$$ \Pi^{\mu \nu}(k)=ie^2\int {d^4p\over(2\pi)^4}
{\rm Tr} \left[ \gamma^\mu {1\over p \llap/-m +i\varepsilon } \gamma^\nu 
{1\over p \llap/-k \llap/-m+i\varepsilon  }\right] 
=\Pi_{(1)}^{\mu \nu}(k)
+\Pi_{(2)}^{\mu \nu}(k)   \eqno{(3.1)} $$
where
$$  \Pi_{(1)}^{\mu \nu}(k)={\alpha \over 2\pi} \left(\Lambda^2+m^2
-{k^2\over 6} \right)g^{\mu \nu}  \eqno{(3.2a)}$$
$$ \Pi_{(2)}^{\mu \nu}(k)={\alpha \over 3\pi}(k^\mu k^\nu- k^2g^{\mu \nu})  
\left[ \ln \left({\Lambda^2\over m^2e}\right) -6\int_0^1 dz z(1-z) 
\ln \left(1-{k^2\over m^2}z(1-z) \right) \right]  \eqno{(3.2b)}  $$
where $\Lambda$ denotes the cutoff momentum. 
Here, the $ \Pi_{(1)}^{\mu \nu}(k)$ corresponds to the quadratic divergence term 
and it has been claimed that this should be discarded since it violates the gauge 
invariance when one considers the counter term of the Lagrangian density. 
The $ \Pi_{(2)}^{\mu \nu}(k)$ term can keep the gauge invariance, and therefore 
one can renormalize it into the new Lagrangian density.

\subsection{Gauge Condition of $\Pi^{\mu \nu}(k)$}
For a long time, people believe that the $\Pi^{\mu \nu}(k)$ should satisfy 
the relation of eq.(1.1)
$$ k_\mu \Pi^{\mu \nu}(k) =0    $$
and this equation is called ``gauge condition of  $\Pi^{\mu \nu}(k)$". 
This is the basic reason why people discarded the first term of eq.(3.1). 
The proof of the above relation seems to be simple and straightforward as 
discussed in the text book of Bjorken and Drell \cite{bd}. Even though they noticed 
that the proof cannot be justified for the infinite integral, they accepted 
the relation as a result in the calculations. 
The serious problem is that, since then, most of the field theory textbooks took 
the gauge condition of $\Pi^{\mu \nu}(k)$ as granted. 

Here, we show that the proof of eq.(1.1) is a simple mistake and the relation 
has no foundation at all. 
First, we present the usual method of the proof, and we rewrite 
the $k_\mu \Pi^{\mu \nu}(k) $ as
$$ k_\mu\Pi^{\mu \nu}(k)=ie^2\int {d^4p\over(2\pi)^4}{\rm Tr}  
\left[ \left({1\over p \llap/-k \llap/-m +i\varepsilon } 
-{1\over p \llap/-m+i\varepsilon  }\right)\gamma^\nu \right] .  \eqno{(3.3)} $$
In the first term, the integration variable should be replaced as
$$ q=p-k $$
and thus one can prove that
$$ k_\mu\Pi^{\mu \nu}(k)=ie^2\int {d^4q\over(2\pi)^4}{\rm Tr}  
\left[ {1\over q \llap/-m +i\varepsilon } 
 \gamma^\nu \right] -
ie^2\int {d^4p\over(2\pi)^4}{\rm Tr}  
\left[ {1\over p \llap/-m+i\varepsilon  }\gamma^\nu \right] =0 .  \eqno{(3.4)} $$
At a glance, this proof looks plausible. However, one can easily notice 
that the replacement of the integration variable is only meaningful when 
the integration is finite. In order to clarify the mathematical mistake in eq.(3.4), 
we present a good example which shows that one cannot make a replacement of 
the integral variable when the integral is infinity. Now, we evaluate 
the following integral
$$ Q =\int_{-\infty}^\infty \left( (x-a)^2-x^2 \right) dx .  \eqno{(3.5)}  $$
Here, if we replace the integral variable in the first term as 
$ x'=x-a$, then we can rewrite eq.(9) as
$$ Q =\int_{-\infty}^\infty \left( {x'}^2dx'-x^2 dx \right) =0 .  \eqno{(3.6)}  $$
However, if we calculate it properly, then we find 
$$ Q =\int_{-\infty}^\infty \left( (x-a)^2-x^2 \right) dx =
\int_{-\infty}^\infty \left( a^2-2ax \right) dx 
 =a^2\times \infty   \eqno{(3.7)}  $$
which disagrees with eq.(3.6). 
If one wishes to carefully calculate eq.(3.5) by replacing the integration 
variable, then one should do as follows
$$ Q = \lim_{\Lambda \rightarrow \infty}  \int_{-\Lambda}^\Lambda 
\left( (x-a)^2-x^2 \right) dx= \lim_{\Lambda \rightarrow \infty}\left[ 
\int_{-\Lambda-a}^{\Lambda-a} {x'}^2  dx'-\int_{-\Lambda}^\Lambda x^2 dx \right] 
= \lim_{\Lambda \rightarrow \infty} 2a^2 \Lambda.  \eqno{(3.8)}  $$
It is clear by now that the replacement of the integration variable in the infinite 
integral should not be made, and this is just the mistake which has been accepted 
as the gauge condition of the $\Pi^{\mu \nu}(k)$ in terms of eq.(1.1). 
Therefore, the requirement of the gauge condition of the vacuum polarization is 
unphysical. In this sense, the calculated result 
of the  $\Pi^{\mu \nu}(k)$  should be the ones which are given in eqs.(3.2). 
Mathematically and physically, there is no other result of the  $\Pi^{\mu \nu}(k)$ 
than eqs.(3.2). 

Some of the readers of Bjorken and Drell's text book \cite{bd} may have a question 
as to why they obtained the vanishing contribution of the $ \Pi_{(1)}^{\mu \nu}(k)$ 
term. For this, one can easily check the defect of the scaling trick in which they 
ignored the scale dependence of the integral ranges, and one sees that the mistake 
in the scaling trick is just similar to the one which is shown in eq.(3.8). 
This clearly shows that one has to be careful for the variable change in the infinite 
integral.

\subsection{Number of Constraints}
If eq.(1.1) were correct, then the number of the constraints due to the gauge 
invariance condition would become four. However, this number is too large 
compared with the condition from the gauge invariance. The gauge fixing 
should be only one condition like Coulomb gauge fixing and so on. 
This clearly indicates that the condition of eq.(1.1) is obviously unphysical. 
The gauge fixing should be done before the field quantization, and there is no additional 
condition of the gauge invariance, apart from eq.(1.3) which is well satisfied 
in the perturbation theory of QED.

\subsection{Dimensional Regularization}
As is well known, the dimensional regularization \cite{thv1,thv2} 
gives the vanishing contribution to the $ \Pi_{(1)}^{\mu \nu}(k)$ term when 
the integration of four dimensions $d^4p$ is replaced by $d^Dp$ with $D=4-\epsilon$.  
However, this regularization is quite odd since 
the result of $ \Pi_{(1)}^{\mu \nu}(k)$ term cannot be recovered when one sets 
$\epsilon \rightarrow 0 $, in contrast to any other regularizations such as 
the $\zeta-$function regularization. For example, in the $\zeta-$function 
regularization, one can, of course, recover the original divergence when one sets 
$\epsilon \rightarrow 0 $. In this respect, the dimensional regularization 
must have some simple mathematical problems which have nothing to do with physics. 
On the other hand, the divergence of the $ \Pi_{(2)}^{\mu \nu}(k)$ term can be 
recovered at the limit of $\epsilon \rightarrow 0$, and this must be connected to the fact 
that the integration in this case is convergent in the complex plane as long as 
one takes the integration dimension of $D=4-\epsilon$ where the contour integration 
at the infinite semi-circle of $R$ vanishes as $R^{-\epsilon}$ with $R\rightarrow 
\infty$. Therefore, the dimensional regularization can only evaluate 
the  $ \Pi_{(2)}^{\mu \nu}(k)$ term properly.  

On the other hand, the reason why the $ \Pi_{(1)}^{\mu \nu}(k)$ term disappears 
in the dimensional regularization can be easily understood as follows. In terms of 
the contour integration at the infinite semi-circle of $R$, it diverges like 
$R^{2-\epsilon}$ with $R\rightarrow \infty$, and therefore it cannot be recovered 
once it is thrown away. Thus, it is now obvious that there is no mathematical reason 
to justify the calculation of the dimensional regularization. 

At this point it may be interesting to clarify the reason as to why the calculated 
result by the dimensional regularization can satisfy the gauge condition of eq.(1.1). 
From eq.(3.3), one can see that the infinite terms are set to zero in the calculation 
of the dimensional regularization while the finite terms can be properly evaluated 
by the replacement of the integral variable, and thus those finite terms of 
the first and the second terms in the right hand side cancel with each other. 
Therefore, the calculation of the dimensional regularization can indeed satisfy 
the gauge condition of eq.(1.1), but this is, of course, accidental.

\subsection{Physical Processes Involving Vacuum Polarizations}
In nature, there are a number of Feynman diagrams which involve the vacuum 
polarization. The best known physical process must be the $\pi^0$ decay into 
two photons, $\pi^0 \rightarrow \gamma +\gamma$. This process of the Feynman 
diagrams can be well calculated in terms of the nucleon and anti-nucleon pair creation 
where these fermions couple to photons \cite{nishi}. In this calculation, one knows that 
the loop integral gives a finite result since the apparent logarithmic divergence 
vanishes to zero due to the kinematical cancellation. Also, the physical process 
of photon-photon scattering involves the box diagrams where electrons and positrons 
are created from the vacuum state. As is well known, the apparent logarithmic 
divergence of this box diagrams vanishes again due to the kinematical cancellation, and 
the evaluation of the Feynman diagrams gives a finite number. This is clear 
since all of the perturbative calculations employ the free fermion basis states 
which always satisfy the current conservation of $ \partial_\mu j^\mu=0 $. 
In these processes, one does not have any additional ``gauge conditions" 
in the evaluation of the Feynman diagrams. In this respect, if the process is physical, 
then the corresponding Feynman diagram should become finite without any further 
constraints of the gauge invariance nor regularizations. 

\section{Effective Lagrangian Density of Heisenberg for Vacuum Polarization}
Here, we briefly review the calculations of the self-energy of photon by Heisenberg and 
Euler \cite{heisen,heisen2}. The calculations are based on the vacuum 
polarization due to the interaction of photon with positive and negative energy fermions. 
In this calculation, they started from the assumption that the negative energy fermions 
can interact with the electromagnetic fields even if they are static fields. This is based 
on the misunderstanding that the negative energy fermions are present as if they were 
the same as the positive energy fermions. In reality, the negative energy fermions 
occupy the negative energy states which are specified by their momenta and there are 
no coordinate dependences in the negative energy fermions. The creation of the fermion 
and anti-fermion pairs can be only possible from the time dependent interactions.

\subsection{Effective Lagrangian Density }
As for the self-energy of photon, they obtain the results which contain the quadratic 
divergence and logarithmic divergence terms. In addition, they obtain some 
finite contributions. Up to this point, their results are practically the same as  
the modern calculations. Now the problem arises when they construct the Lagrangian 
density from their calculated results. They write the effective Lagrangian density 
$$ {\cal L}={1\over 2} (\bm{E}^2-\bm{B}^2) +\alpha \int_0^\infty {d\eta\over \eta^3} 
e^{-\eta} \left\{ i\eta^2 (\bm{E}\cdot \bm{B}) {\cos( {\eta\over {\cal E}_0}
\sqrt{\bm{E}^2-\bm{B}^2+2i\bm{E}\cdot \bm{B}}) + c.c. \over 
\cos( {\eta\over {\cal E}_0}\sqrt{\bm{E}^2-\bm{B}^2+2i\bm{E}\cdot \bm{B}})-c.c.} 
+{\cal E}_0^2-{\eta^2\over 3}(\bm{E}^2-\bm{B}^2) \right\}  \eqno{(4.1)}  $$
where ${\cal E}_0$ is given as ${\cal E}_0={m^2\over e}$. This effective 
Lagrangian density formulation is a dangerous attempt since eventually those fields 
should be treated as field operators after the field quantization. We believe that any 
fields which are quantized should be fundamental fields, and these fields which appear 
in Lagrangian density of eq.(4.1) are not the fundamental fields any more. 
In this case, we may ask as to what it means by the effective Lagrangian density ? 
The basic question is the physical meaning of the fields $\bm{E}$ and $\bm{B}$ 
since they interact with each other by themselves. In particular, one cannot define 
the free electromagnetic field from this Lagrangian density since it corresponds to 
the physical state with no matter fields. This is very serious, and in fact, it is not 
consistent with observations that photon is always in a free state. 

In summary, the problem of the effective Lagrangian method is that it is 
inconsistent with the definition of the vacuum state and therefore it cannot 
agree with the renormalization scheme.  

\subsection{Proper Treatment and Renormalization Scheme}
By now it becomes clear that the treatment of Heisenberg concerning the self-energy 
of photon is incorrect. From the physical quantity which is calculated in the second order 
perturbation theory of the vacuum polarization due to photon, one can obtain 
the energy of the process as the function of the four momentum $k^\mu $ of photon
$$ \Delta E= f(k^\mu) . \eqno{(4.2)} $$
What one should do or one can do is that one should find the same shape as the original 
Lagrangian density as given
$$ {\cal L}_0={1\over 2} (\bm{E}^2-\bm{B}^2)=-{1\over 4} F_{\mu \nu} 
F^{\mu \nu} \eqno{(4.3)} $$
by redefining the fields $\bm{E}$ and $\bm{B}$. 
In this respect, one sees that the only possible shape of the vacuum polarization 
energy should be written as 
$$ \Delta E= C_0 k^2  \eqno{(4.4)} $$
where $C_0$ denotes some constant which may well be an infinite quantity. 
This is clear since, by noting $\epsilon_\mu k^\mu=0$ and $\epsilon_\mu \epsilon^\mu =1$, 
one can rewrite the above energy  as 
$$ \Delta E= C_0 \epsilon^\mu( g_{\mu \nu}k^2-k_\mu k_\nu)\epsilon^\nu  \eqno{(4.5)} $$
which corresponds to the following Lagrangian density 
$$ {\cal L}'=C_0 A^\mu(g_{\mu \nu}\partial^2- \partial_\mu \partial_\nu ) 
A^\nu . \eqno{(4.6)} $$
This can be renormalized into the original Lagrangian density. However, in reality, 
the calculated result of the finite term is written by eq.(4.1), but in addition, 
there are other infinite terms present in the vacuum polarization contributions. 
The calculated result of the vacuum polarization energy can be given as
$$ \Delta E= {\alpha \over 2\pi} \left(\Lambda^2+m^2\right) -
{\alpha \over 3\pi} \left({1\over 4}+C' \right) k^2  \eqno{(4.7)} $$
where $C'$ is written as
$$ C'= \ln \left({\Lambda^2\over m^2e}\right) -6\int_0^1 dz z(1-z) 
\ln \left(1-{k^2\over m^2}z(1-z) \right)  . $$
Therefore, there is no way to renormalize the self-energy of photon into the original 
Lagrangian density, and the only physically correct way is to discard all of 
the self-energy of photon contributions. This conclusion is just the same as the one 
that is presented in section 2 in terms of Tomonaga's conjecture. 
The renormalization procedure without the photon self-energy effect is most natural, and 
up to the present stage, there is no experimental evidence that shows any of the effects 
arising from the self-energy of photon.

\subsection{Improper Application of Effective Lagrangian}
For a long time, people have been applying the effective Lagrangian density of Heisenberg 
to physical processes, and some of the applications give rise to rather serious problems 
in the fundamental interactionsD

\subsubsection{Uehling Potential}
The most important of all is the Uehling potential which is obtained by Uehling 
\cite{ueh}.  By now it is clear that there is no modification of the Coulomb potential, 
and there is no way to obtain the Uehling potential. This Uehling potential is obtained 
by making double mistakes. The first one is just related to the renormalization 
procedure, and the effective Lagrangian method gives rise to the unphysical effect 
as discussed above. The other important mistake is connected to the Coulomb potential. 
As can be easily seen from the field quantization procedure, one has to first fix 
the gauge, and by choosing the Coulomb gauge fixing, one can obtain the equation 
for the $A_0$ field which becomes a constraint equation. Therefore, one finds that 
the $A_0$ field becomes time independent and thus  the $A_0$ field can be solved 
exactly as given in eq.(2.12). There is no modification of the Coulomb field, and 
this result is stronger than the renormalization procedure. This means that the $A_0$ 
field should not be quantized, and therefore it is not involved in the renormalization 
scheme from the beginning. 

\subsubsection{Photon-Photon Scattering}
Further, there is some application of the effective Lagrangian density to the 
evaluation of the scattering cross section between two photons \cite{karp,landau}. 
The cross section which is derived by making use of the effective  Lagrangian density 
has no foundation at all, and it is physically incorrect. The proper cross section 
between two photons has been calculated, and it is completely different from 
the old version of the photon-photon cross section \cite{kanda}. 
In fact, the photon-photon cross section is rather large, and it should be well 
detectable if one can control the time scale of photon in order to make a head-on 
collisions. However, the beam focusing of photon at low energy may well be difficult, 
and it should be non-trivial to make the focusing of photon 
in a sufficiently high standard \cite{fujita3}. 

\section{Conclusions}
We have examined Tomonaga's conjecture that the self-energy of photon should vanish 
to zero. Here, we prove that his conjecture is correct, and the self-energy of photon 
should indeed be taken to be zero. This proof is mainly based on the exact 
Lippmann-Schwinger equation for the vector potential which clearly shows that there is 
no wave function renormalization possible for any iteration procedures. In addition, 
the photon self-energy violates the Lorentz invariance, and it is unphysical. Therefore 
it is simply discarded so as to keep the basic symmetry property of the system.  

In addition, we critically review the photon self-energy treatment which has been 
considered to be a standard procedure until now. It is known that 
the quadratic divergence of the photon self-energy diagram is the only defect 
in the renormalization scheme of QED, and people discarded this divergent parts 
by requiring the gauge condition of the vacuum polarization tensor $\Pi^{\mu \nu}(k)$ 
since it is believed that $ k_\mu \Pi^{\mu \nu}(k) =0 $ should hold. 
This condition of the gauge invariance has been used 
in the calculations of many vacuum polarization diagrams. 
However, we prove that this relation does not hold since it is derived by 
the simple mathematical mistake. Even though some people noticed that the derivation 
is wrong in mathematics, they continue to employ this gauge condition of the vacuum 
polarization tensor in their calculations of Feynman diagrams. Therefore, there are 
quite a few examples of the Feynman diagram evaluations which are incorrect because of 
the wrong condition of the gauge invariance. 

By now, it becomes clear that photon should always stay massless, and 
there is no renormalization of photon propagation. It is just simple. The only procedure 
one should consider is the self-energy of fermion, and there is neither conceptual nor 
technical difficulty of treating the renormalization of fermion self-energy.

\vspace{1cm}
We are grateful to Professor M. Sato for the information on the discussions of the 
renormalization scheme he had with S. Tomonaga.


\end{document}